\documentclass[pdflatex,sn-mathphys-num]{sn-jnl}
\usepackage{amsmath,amssymb,bm,braket}
\usepackage{booktabs,tabularx,array}
\usepackage{graphicx}
\usepackage{quantikz}

\usepackage[nolinenumbers,noshowlabels]{jadeleon} % {{{*
\newcommand{\cE}{\mathcal{E}}
\newcommand{\id}{\mathbb{I}}
\newcommand{\pstar}{p^{\ast}}
\theoremstyle{thmstyleone}

\begin{document}

\title[Local and correlated Pauli noise in teleportation]{Characterizing Local and Correlated Pauli Noise in Quantum Teleportation Circuits}

\author*[1]{\fnm{L.} \sur{Anjos}}
\email{laisajsjesus@gmail.com}

\author[2]{Jose Alfredo de Leon}
\author[1]{\fnm{R. S.} \sur{Sarthour}}
\author[3]{\fnm{Alejandro} \sur{Fonseca}}

\affil*[1]{\orgname{Centro Brasileiro de Pesquisas F\'isicas (CBPF)},
\orgaddress{\city{Rio de Janeiro}, \state{RJ}, \country{Brazil}}}

\affil[2]{\ifunam}

\affil[3]{\orgdiv{Departamento de F\'isica},
\orgname{Universidad Nacional de Colombia},
\orgaddress{\country{111321 Bogotá, Colombia}}}

\abstract{We analytically characterize how the location and the local or correlated structure of stochastic Pauli errors affect a three-qubit coherent quantum-teleportation circuit. Independent local channels and a fully correlated two-qubit channel are inserted separately in six configurations. Propagating each Pauli-error realization through the Clifford circuit yields the effective channel on Bob's qubit and analytical expressions for the average and state-resolved fidelities. Under a common ordered-pair probability convention, the six configurations form three pairs that induce equivalent reduced channels on Bob. Within each pair, corresponding Pauli-error realizations have the same probabilities and produce the same effective Pauli operator on Bob's output state. The equivalence therefore holds for every input state within the stochastic Pauli models considered here, rather than only for the average fidelity. Correlated $Z$ errors do not affect the output for the first pair; local $X$ and $Z$ errors exhibit complementary qubit sensitivity for the second; and correlated same-axis Pauli errors cancel for the third. These results classify the error response of the selected circuit configurations and show that it depends jointly on the error statistics, Pauli axis, and position within the protocol.}


\maketitle

\section{Introduction}
One of the main goals of quantum information processing is to exploit quantum resources such as coherence, interference, and entanglement to perform computational and communication tasks that have no direct classical equivalent \cite{NielsenChuang}. Quantum teleportation is one of the foundational protocols of quantum information science: an unknown quantum state is transferred from Alice to a physically separated receiver, Bob, by combining a shared entangled pair of qubits with two bits of classical information, without physically transmitting the input system itself \cite{Bennett1993}. Beyond its foundational relevance, teleportation is a basic primitive for quantum communication and quantum-network architectures \cite{Pirandola2016}. Experimental implementations have been demonstrated using different physical systems, including photons and atomic ensembles \cite{bouwmeester1997experimental, ExpTele2, ren2017ground,TeleAtomicEnsemble}.

In any physical implementation, interactions with uncontrolled degrees of freedom and imperfections in state preparation, gates, and measurements alter the intended evolution \cite{BreuerPetruccione,Zurek2003}. These effects are particularly relevant for noisy intermediate-scale quantum devices, whose performance is limited by imperfect operations and finite coherence times \cite{Preskill2018,Kandala2019}. Quantum error-correcting codes can suppress faults when sufficiently large and well-controlled devices are available \cite{Shor1995,Steane1996}. For near-term protocols, however, it is also essential to determine which errors reach the output, how they propagate through the circuit, and which qubits or circuit regions should be prioritized for protection.

Teleportation through a noisy entangled resource can itself be represented as an effective quantum channel \cite{Bowen2001,Horodecki1999}. Previous studies have investigated noisy entangled resources and recovery operations, Pauli errors applied at different stages, beneficial effects of noise, and correlations or memory in teleportation channels \cite{Oh2002,Yeo_2009,Fortes2015,Fortes2016,articleLi,Macchiavello2002,Dennis2005,Cunha2017,Fonseca2019}. These works establish several of the mechanisms used below, including the dependence of fidelity on the noise position and axis and the cancellation of particular Pauli-error combinations. Here we address a more specific circuit-level question: whether errors inserted on different qubit pairs and at different positions induce the same complete reduced channel on Bob. This requires comparing the event-by-event effective Pauli maps, rather than only their average fidelities.

In this work, we analyze stochastic Pauli errors along the \(X\), \(Y\), and \(Z\) axes separately in six selected configurations, denoted by \(T_1,\ldots,T_6\), of a three-qubit coherent teleportation circuit. We compare statistically independent local errors with fully correlated two-qubit error events and evaluate the reduced channel induced on Bob's qubit. Because the teleportation circuit is a Clifford circuit, each Pauli-error realization can be propagated analytically through the remaining gates. This procedure determines which Pauli operator effectively acts on Bob's output state, from which the average and state-resolved fidelities are calculated. The principal contribution is the explicit classification of the six configurations into pairs with identical reduced channels under the stated event and probability correspondence.

The analysis reveals three error-response classes. The six configurations form the channel-equivalent pairs \(T_1\leftrightarrow T_6\), \(T_2\leftrightarrow T_5\), and \(T_3\leftrightarrow T_4\). Correlated \(Z\) errors do not affect Bob's output at \(T_1/T_6\); local \(X\) and \(Z\) errors show complementary qubit sensitivity at \(T_2/T_5\); and correlated same-axis Pauli errors cancel at \(T_3/T_4\). We derive analytical expressions for the corresponding average fidelities and identify the input-state axes preserved by each effective channel. Section~2 introduces the coherent protocol, the noise models, and the fidelity measure. Section~3 defines the error locations and the analytical procedure. Section~4 presents the three response classes and demonstrates their effective-channel symmetry. Section~5 summarizes the conclusions and perspectives, while detailed calculations are provided in the Supplementary Information.

\section{Teleportation protocol, Pauli Noise, and Fidelity}

\subsection{Quantum teleportation protocol}
Quantum teleportation, originally introduced by Bennett \textit{et al.} \cite{Bennett1993}, enables the transfer of an unknown quantum state between two spatially separated parties (Alice and Bob) using a shared entangled state, local operations, and classical communication \cite{NielsenChuang}.

Let qubit $0$ (belonging to Alice) hold the arbitrary single-qubit target state to be teleported:
\begin{equation}
 \ket{\psi}_0 = \cos\left(\frac{\theta}{2}\right)\ket{0}_0 + e^{i\phi}\sin\left(\frac{\theta}{2}\right)\ket{1}_0,
 \label{eq:inputstate}
\end{equation}
where $\theta \in [0, \pi]$ and $\phi \in [0, 2\pi)$ parametrize the single-qubit state space on the Bloch sphere. Qubits $1$ and $2$, held by Alice and Bob respectively, are initialized in the product state $\ket{00}_{12}$.

\subsubsection{Entangled-resource preparation}
First, a maximally entangled state shared between Alice and Bob is prepared. Applying a Hadamard gate $H_1$ followed by a controlled-NOT gate $\mathrm{CX}_{1\to2}$ creates the maximally entangled Bell state $\ket{\Phi^+}_{12} = \frac{1}{\sqrt{2}}(\ket{00}_{12} + \ket{11}_{12})$. The combined state of the three-qubit system in $\mathcal{H}_0 \otimes \mathcal{H}_1 \otimes \mathcal{H}_2$ prior to teleportation is described by the global product state:
\begin{equation}
\ket{\Psi_0}_{012} = \ket{\psi}_0 \otimes \ket{\Phi^+}_{12}.
\label{eq:initial_joint_state}
\end{equation}

In the standard measurement-based protocol, Alice performs a Bell-basis measurement on qubits \(0\) and \(1\). Depending on the two-bit measurement outcome, Bob's qubit is related to the input state by one of the Pauli operators \(I\), \(X\), \(Z\), or \(XZ\). Alice communicates the result to Bob, who applies the corresponding correction. Before receiving this classical information, Bob's reduced state is maximally mixed, $\rho_B = \frac{1}{2}I_2$.

\subsubsection{Coherent circuit implementation}
Using a coherent implementation of the teleportation protocol, the intermediate measurements and classically controlled corrections are replaced by quantum-controlled unitary operations. Rather than performing non-unitary projective measurements followed by classically controlled corrections, Alice's basis transformation and Bob's conditional corrections are executed using quantum entangling gates.

Specifically, Alice applies $\mathrm{CX}_{0\to1}$ followed by $H_0$. This unitary transformation rotates the Bell basis of qubits $0$ and $1$ back into the computational basis $\{|00\rangle, |01\rangle, |10\rangle, |11\rangle\}$. At this intermediate stage, the joint state evolves to:
\begin{equation}
\begin{aligned}
(H_0\mathrm{CX}_{0\to1}\otimes I_2)\ket{\Psi_0}_{012}
=
\frac{1}{2}\big(
&\ket{00}_{01}\ket{\psi}_2
+\ket{01}_{01}X_2\ket{\psi}_2\\
&+\ket{10}_{01}Z_2\ket{\psi}_2
+\ket{11}_{01}X_2Z_2\ket{\psi}_2
\big).
\end{aligned}
\end{equation}

To complete the protocol coherently, Bob's conditional corrections are implemented via entangling gates controlled by Alice's qubits: a controlled-NOT gate $\mathrm{CX}_{1\to2}$ and a controlled-$Z$ gate $\mathrm{CZ}_{0,2}$. The full unitary operator describing the circuit protocol is:
\begin{equation}
U_{\rm tel} = \mathrm{CZ}_{0,2}\,\mathrm{CX}_{1\to2}\,H_0\,\mathrm{CX}_{0\to1}\,\mathrm{CX}_{1\to2}\,H_1,
\label{eq:utel}
\end{equation}
where gates act sequentially from right to left. Under ideal (noiseless) conditions, the action of $U_{\rm tel}$ on the initial state simplifies to:
\begin{equation}
U_{\rm tel} \left(\ket{\psi}_0\ket{00}_{12}\right) = \ket{+}_0\ket{+}_1\ket{\psi}_2,
\label{eq:idealmap}
\end{equation}
where $\ket{+} = (\ket{0}+\ket{1})/\sqrt{2}$. Equation~\eqref{eq:idealmap} explicitly demonstrates the complete factorization between Alice's local subsystem and Bob's target qubit. Tracing out Alice's qubits yields Bob's ideal reduced density matrix:
\begin{equation}
\rho_B^{\rm id} = \operatorname{Tr}_{01}\!\left[ U_{\rm tel} \left(\ket{\psi}\!\bra{\psi}_0 \otimes \ket{00}\!\bra{00}_{12}\right) U_{\rm tel}^{\dagger} \right] = \ket{\psi}\!\bra{\psi}_2.
\label{eq:idealbob}
\end{equation}

\subsubsection{Analytical advantages for open quantum systems}
For ideal operations, measuring Alice's qubits in the computational basis and applying the usual classically controlled corrections gives the same reduced output state on Bob as the coherent circuit above. The coherent formulation is used here as an analytical representation of this ideal map; the cross-party controlled gates should not be interpreted as an LOCC implementation or as a detailed hardware model of measurement and feed-forward errors. Its advantage is that the complete protocol is unitary and Clifford, so the stochastic Pauli errors considered here remain Pauli operators when propagated through the subsequent gates. This property is used to determine Bob's effective channel analytically.

\subsection{Local and correlated Pauli channels}
In the context of open quantum systems, quantum channels serve as the foundational mathematical framework to describe non-unitary physical dynamics, such as quantum noise, decoherence, and energy dissipation resulting from interactions with an external environment \cite{NielsenChuang,BreuerPetruccione}. Physically, a quantum channel represents the effective evolution of a principal system coupled to an environment—typically assumed to be initially uncorrelated—after tracing out the environmental degrees of freedom. Formally, a quantum channel is defined as a completely positive trace-preserving (CPTP) linear map \(\cE:\mathcal{L}(\mathcal{H})\rightarrow\mathcal{L}(\mathcal{H})\). Linearity guarantees that convex combinations of density matrices are mapped into convex combinations of evolved states, while trace preservation ensures that \(\operatorname{Tr}[\cE(\rho)]=\operatorname{Tr}(\rho)=1\), conserving total probability. Complete positivity is necessary to guarantee that the evolved state remains positive semi-definite even when the system is entangled with an arbitrary ancillary system. In the Kraus representation (or operator-sum representation), the action of the channel is written as
\begin{equation}
\cE(\rho)=\sum_k K_k\rho K_k^\dagger,
\qquad
\sum_k K_k^\dagger K_k=\id,
\label{eq:krausgeneral}
\end{equation}
where \(\{K_k\}\) is a set of Kraus operators acting on the system's Hilbert space \(\mathcal{H}\). These operators encapsulate the operational effects of the system-environment interaction without requiring explicit knowledge of the environment's internal dynamics.

A single-qubit Pauli channel arises when the Kraus operators are constructed from the Pauli operators $\{\id, X, Y, Z\}$. To analyze the effect of each Pauli axis separately, we consider the single-axis Pauli channel
\begin{equation}
\cE_{P,p}(\rho)
=
(1-p)\rho+pP\rho P,
\qquad
P\in\{X,Y,Z\},
\label{eq:singlepauli}
\end{equation}
where \(p\in[0,1]\) is the probability that the Pauli error \(P\) occurs. The channel has the Kraus operators \(K_0=\sqrt{1-p}\,\id\) and \(K_1=\sqrt{p}\,P\).

Following the Pauli-component description of quantum channels discussed in Ref.~\cite{DeLeon2022}, an arbitrary \(N\)-qubit density matrix can be expanded in the basis of Pauli strings as
\begin{equation}
\rho
=
\frac{1}{2^N}
\sum_{\vec{\alpha}}
r_{\vec{\alpha}}\,
\sigma_{\vec{\alpha}},
\qquad
\sigma_{\vec{\alpha}}
=
\sigma_{\alpha_1}\otimes\cdots\otimes\sigma_{\alpha_N},
\label{eq:pauliexpansion}
\end{equation}
where \(r_{\vec{\alpha}} = \operatorname{Tr}(\rho\,\sigma_{\vec{\alpha}})\) are the corresponding Pauli components. The channel in Eq.~\eqref{eq:singlepauli} leaves components that commute with \(P\) unchanged and multiplies components that anticommute with \(P\) by \(1-2p\). This Pauli-string representation is also the basis used below to track each error realization through the Clifford circuit.

We compare two idealized error-statistics models: independent local errors and fully correlated same-axis errors:

\subsubsection{Local independent errors}

For qubits \(a\) and \(b\), the local noise model is defined as the tensor
product of two statistically independent single-qubit channels:
\begin{equation}
\cE^{\rm loc}_{P;p_a,p_b}
=
\cE^{(a)}_{P,p_a}\otimes\cE^{(b)}_{P,p_b}.
\label{eq:local}
\end{equation}
Its action is
\begin{align}
\cE^{\rm loc}_{P;p_a,p_b}(\rho)
={}&(1-p_a)(1-p_b)\rho
+p_a(1-p_b)P_a\rho P_a \nonumber\\
&+(1-p_a)p_bP_b\rho P_b
+p_ap_b(P_aP_b)\rho(P_aP_b),
\label{eq:localexpanded}
\end{align}
where \(P_a\) and \(P_b\) occur independently with probabilities \(p_a\) and \(p_b\), respectively.

\subsubsection{Correlated errors}

For comparison, we consider a fully correlated model in which the same Pauli error acts simultaneously on both qubits:
\begin{equation}
\cE^{\rm corr}_{P,p}(\rho)
=
(1-p)\rho
+p(P_aP_b)\rho(P_aP_b).
\label{eq:corr}
\end{equation}
The corresponding Kraus operators are \(K_0=\sqrt{1-p}\,\id_{ab}\) and \(K_1=\sqrt{p}\,(P_a\otimes P_b)\). Thus, with probability \(1-p\), no error occurs, whereas with probability \(p\), the joint error \(P_a\otimes P_b\) acts on the two qubits. This is an idealized fully correlated stochastic Pauli model, which may represent two errors driven by a common classical fluctuation. It does not by itself establish quantum environmental correlations, non-Markovianity, or temporal memory. Comparing it with the independent model isolates how the joint-event structure changes the logical error reaching Bob. A direct comparison of correlation at fixed single-qubit marginals is obtained by setting \(p_a=p_b=p\) in the local model.

\subsection{Teleportation fidelity}
The performance of the teleportation protocol is quantified by the fidelity between the input state \(\ket{\psi}\) and Bob's output state \(\rho_B\) \cite{Jozsa1994,NielsenChuang}. For arbitrary density operators \(\rho\) and \(\sigma\), the Uhlmann--Jozsa fidelity is defined as
\begin{equation}
F(\rho,\sigma)
=
\left[
\operatorname{Tr}
\sqrt{\sqrt{\rho}\,\sigma\sqrt{\rho}}
\right]^2.
\label{eq:fidelitygeneral}
\end{equation}
In the teleportation scenario, after tracing out Alice's qubits from the joint three-qubit system, Bob holds the reduced density matrix $\rho_B(\psi) = \operatorname{Tr}_{01}[\mathcal{E}(\ket{\psi}\!\bra{\psi}_0 \otimes \ket{00}\!\bra{00}_{12})]$, where $\mathcal{E}$ denotes the complete noisy quantum channel mapping the circuit. Since the target state $\ket{\psi}$ is pure, Eq.~\eqref{eq:fidelitygeneral} simplifies to the state overlap:
\begin{equation}
 F(\psi) = \bra{\psi}\rho_B(\psi)\ket{\psi}.
 \label{eq:statefidelity}
\end{equation}

\subsubsection{Average fidelity}

Since the single-state fidelity $F(\psi)$ strongly depends on the specific orientation of the target state on the Bloch sphere—often exhibiting anisotropic behavior depending on whether $\ket{\psi}$ aligns with the dominant noise axes—a single state evaluation cannot capture the overall performance of the quantum channel \cite{Fonseca2019}. To establish an operational, state-independent benchmark that quantifies the protocol's capacity to transmit unknown or arbitrary quantum information without prior knowledge of the input, we evaluate the average fidelity uniformly over the entire single-qubit state space. Using the Bloch sphere parametrization for a general pure target state,
\[
\ket{\psi(\theta,\phi)}
=
\cos\left(\frac{\theta}{2}\right)\ket{0}
+
e^{i\phi}
\sin\left(\frac{\theta}{2}\right)\ket{1},
\]
the state-averaged fidelity $\overline{F}$ is formally defined via continuous integration over the uniform Haar measure (corresponding to the normalized Fubini-Study volume element):
\begin{equation}
\overline{F}
=
\frac{1}{4\pi}
\int_0^{2\pi}\!d\phi
\int_0^\pi \!d\theta\,
\sin\theta\,
\bra{\psi(\theta,\phi)}\rho_B(\theta,\phi)\ket{\psi(\theta,\phi)}.
\label{eq:haaraverage}
\end{equation}
This uniform Haar integration ensures that the resulting measure is invariant under arbitrary unitary basis changes, providing an unbiased global metric that can be directly benchmarked against classical state-transfer thresholds.

For the channels considered here, the fidelity can be written as
\[
F(\psi)
=
\operatorname{Tr}
\left[
\rho_\psi\,\mathcal{E}(\rho_\psi)
\right],
\qquad
\rho_\psi=\ket{\psi}\!\bra{\psi},
\]
which is quadratic in the input-state projector \(\rho_\psi\). Therefore, its uniform average can be evaluated using a quantum-state 2-design. For a single qubit, the six eigenstates of the Pauli operators form such a 2-design:
\begin{equation}
\mathbb{S}_6
=
\left\{
\ket{0},
\ket{1},
\ket{+x},
\ket{-x},
\ket{+y},
\ket{-y}
\right\},
\label{eq:sixstates}
\end{equation}
where
\(\ket{\pm x}=(\ket{0}\pm\ket{1})/\sqrt{2}\) and
\(\ket{\pm y}=(\ket{0}\pm i\ket{1})/\sqrt{2}\).
Consequently, the uniform state average is obtained from the discrete average
\begin{equation}
\overline{F}
=
\frac{1}{6}
\sum_{\ket{\psi}\in\mathbb{S}_6}
F(\psi).
\label{eq:sixaverage}
\end{equation}
This six-state average is equal to the Haar average for the fidelity considered here \cite{Bowdrey2002,NIELSEN2002249}.

In addition, note that in the absence of shared entanglement, a measure-and-prepare strategy—where Alice measures a single copy of a pure input state uniformly distributed over the Bloch sphere and transmits the classical result to Bob—can achieve an average fidelity of at most $2/3$. This threshold is known as the classical fidelity \cite{Massar1995}.

\subsubsection{Effective single-axis Pauli channel}

Consider the effective channel on Bob's qubit,
\begin{equation}
\mathcal{E}_{Q,r}(\rho)
=
(1-r)\rho+rQ\rho Q,
\qquad
Q\in\{X,Y,Z\},
\label{eq:effectivepaulichannel}
\end{equation}
where \(r\) is the effective probability that the Pauli error \(Q\) acts
on Bob's output state. Substitution into Eq.~\eqref{eq:sixaverage} gives
\begin{equation}
\overline{F}
=
1-\frac{2r}{3}.
\label{eq:paulifid}
\end{equation}
The origin of Eq.~\eqref{eq:paulifid} can be understood directly from the six-state set. The two eigenstates of \(Q\), located along the corresponding Bloch-sphere axis, are preserved by the channel and have fidelity \(F=1\). For each of the four states lying on the two orthogonal axes, the Pauli operator \(Q\) maps the input state to an orthogonal state, resulting in \(F=1-r\). Therefore, identifying the effective Pauli operator on Bob's qubit determines both the average fidelity and the input-state axis preserved by the channel.

%%%%%%%%%%%%%%%%%%%%%%%%%%%%%%%%%%%%%%%%%%%%%%%%%%%%%%%%%%%%%%%%%%%%%%%%%%%%%

\section{Circuit Architecture and Noise Modeling}
\subsection{Noise configurations}
To systematically analyze how the effect of Pauli noise depends on its location in the coherent teleportation circuit, we divide the protocol into three operational stages:
\begin{enumerate}
    \item \emph{Entangled-resource preparation}: preparation of the Bell  state on qubits \(1\) and \(2\) using \(H_1\) and  \(\mathrm{CX}_{1\to2}\).

    \item \emph{Bell-basis transformation}: application of  \(\mathrm{CX}_{0\to1}\) and \(H_0\) to Alice's qubits \(0\) and \(1\).

    \item \emph{Coherent corrections}: application of  \(\mathrm{CX}_{1\to2}\) and \(\mathrm{CZ}_{0,2}\), controlled by  Alice's qubits and acting on Bob's qubit \(2\).
\end{enumerate}

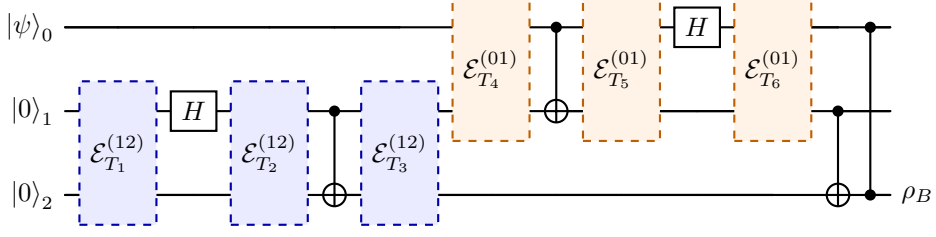
\begin{figure*}[t]
\centering
\begin{quantikz}[row sep=0.32cm,column sep=0.19cm]
\lstick{$\ket{\psi}_0$}
 & \qw
 & \qw
 & \qw
 & \qw
 & \qw
 & \gate[2,style={dashed,fill=orange!10,draw=orange!75!black}]{\cE_{T_4}^{(01)}}
 & \ctrl{1}
 & \gate[2,style={dashed,fill=orange!10,draw=orange!75!black}]{\cE_{T_5}^{(01)}}
 & \gate{H}
 & \gate[2,style={dashed,fill=orange!10,draw=orange!75!black}]{\cE_{T_6}^{(01)}}
 & \qw
 & \ctrl{2}
 & \qw \\
\lstick{$\ket{0}_1$}
 & \gate[2,style={dashed,fill=blue!8,draw=blue!65!black}]{\cE_{T_1}^{(12)}}
 & \gate{H}
 & \gate[2,style={dashed,fill=blue!8,draw=blue!65!black}]{\cE_{T_2}^{(12)}}
 & \ctrl{1}
 & \gate[2,style={dashed,fill=blue!8,draw=blue!65!black}]{\cE_{T_3}^{(12)}}
 & \qw
 & \targ{}
 & \qw
 & \qw
 & \qw
 & \ctrl{1}
 & \qw
 & \qw \\
\lstick{$\ket{0}_2$}
 & \qw
 & \qw
 & \qw
 & \targ{}
 & \qw
 & \qw
 & \qw
 & \qw
 & \qw
 & \qw
 & \targ{}
 & \control{}
 & \rstick{$\rho_B$}\qw
\end{quantikz}
\caption{Coherent quantum teleportation circuit with the six noise insertion locations \(T_1\)--\(T_6\). The blue boxes \(T_1\)--\(T_3\) represent noise acting on the resource pair \((1,2)\), whereas the orange boxes \(T_4\)--\(T_6\), noise acting on Alice's pair \((0,1)\). Only one noise location is considered in each calculation. The configurations \(T_3\) and \(T_4\) correspond to the same circuit boundary but act on different qubit pairs.}
\label{fig:types}
\end{figure*}

Figure~\ref{fig:types} shows the six noise configurations, denoted by \(T_1,\ldots,T_6\), placed at five temporal circuit boundaries. They were selected to compare the resource pair \((1,2)\) during its preparation with Alice's pair \((0,1)\) during the Bell-basis transformation and before the coherent corrections. They are not intended as an exhaustive list of every possible qubit pair at every circuit boundary. In each analytical calculation, the noise channel is inserted in only one configuration.

As summarized in Table~\ref{tab:types}, the six configurations are divided according to the qubit pair on which the noise channel acts:
\begin{itemize}
    \item \textbf{Resource-pair configurations (\(T_1\)--\(T_3\)):} the noise channel acts on qubits \(1\) and \(2\), before, during, or immediately after the preparation of the entangled resource.

    \item \textbf{Alice-pair configurations (\(T_4\)--\(T_6\)):} the noise channel acts on qubits \(0\) and \(1\) at different stages of the Bell-basis transformation and before the coherent corrections.
\end{itemize}

Configurations \(T_3\) and \(T_4\) are located at the same circuit boundary, after the preparation of the Bell pair and before \(\mathrm{CX}_{0\to1}\), but act on different qubit pairs. Configuration \(T_3\) acts on the resource pair \((1,2)\), whereas \(T_4\) acts on Alice's pair \((0,1)\). Configuration \(T_6\) is located after \(H_0\) and before the coherent correction gates. These placements allow us to compare the reduced channels induced on Bob by errors acting on different qubits and at different stages of the protocol.

\begin{table}[b]
\caption{Noise insertion locations in chronological circuit order.}
\label{tab:types}
\centering
\small
\begin{tabularx}{\columnwidth}{c c X}
\toprule
Configuration & Qubit pair & Insertion location \\
\midrule
\(T_1\) & \((1,2)\) & Before \(H_1\) \\
\(T_2\) & \((1,2)\) & Between \(H_1\) and \(\mathrm{CX}_{1\to2}\) \\
\(T_3\) & \((1,2)\) & Between \(\mathrm{CX}_{1\to2}\) and \(\mathrm{CX}_{0\to1}\) \\
\(T_4\) & \((0,1)\) & Between \(\mathrm{CX}_{1\to2}\) and \(\mathrm{CX}_{0\to1}\) \\
\(T_5\) & \((0,1)\) & Between \(\mathrm{CX}_{0\to1}\) and \(H_0\) \\
\(T_6\) & \((0,1)\) & Between \(H_0\) and the coherent corrections \\
\bottomrule
\end{tabularx}
\end{table}

\subsection{Error-probability convention}

For each ordered qubit pair \((a,b)\) listed in Table~\ref{tab:types}, we assign the single-qubit error probabilities as
\begin{equation}
p_a=p,
\qquad
p_b=\pstar.
\label{eq:probabilityconvention}
\end{equation}
Therefore, for the resource-pair configurations \(T_1\)--\(T_3\), which act on the ordered pair \((1,2)\), the error probabilities are \(p_1=p\) and \(p_2=\pstar\). For the Alice-pair configurations \(T_4\)--\(T_6\), which act on the ordered pair \((0,1)\), they are \(p_0=p\) and \(p_1=\pstar\).

This convention allows the two qubits to have different error probabilities and assigns the probabilities consistently to the first and second members of each ordered pair. Consequently, the paired-channel equalities derived below are conditional on this ordered-pair relabeling; they do not state that moving a fixed hardware noise source between physical qubits preserves its rate. The analytical expressions are valid for arbitrary \(p,\pstar\in[0,1]\). In the plots, one probability is varied while the other is fixed to show unequal local rates. The superscript in \(\pstar\) only distinguishes the second probability and does not denote complex conjugation.

\subsection{Analytical evaluation via Pauli propagation}
For each noise location \(T\in\{T_1,\ldots,T_6\}\), we decompose the noiseless teleportation unitary as
\begin{equation}
U_{\rm tel}=W_TV_T,
\label{eq:circuitpartition}
\end{equation}
where \(V_T\) contains the gates applied before the noise location and \(W_T\) contains the gates applied after it. For the global input state
\[
\rho_{\rm in}
=
\ket{\psi}\!\bra{\psi}_0
\otimes
\ket{00}\!\bra{00}_{12},
\]
the final three-qubit state is
\begin{equation}
\rho_f^{(T)}
=
W_T\,
\cE_T
\left(
V_T\rho_{\rm in}V_T^\dagger
\right)
W_T^\dagger.
\label{eq:noisyevolution}
\end{equation}
Bob's reduced output state is then
\begin{equation}
\rho_B^{(T)}
=
\operatorname{Tr}_{01}
\left[
\rho_f^{(T)}
\right].
\label{eq:reduced_bob}
\end{equation}
At each location, \(\cE_T\) is either the local channel \(\cE^{\rm loc}_{P;p,\pstar}\) in Eq.~\eqref{eq:local} or the correlated channel \(\cE^{\rm corr}_{P,p}\) in Eq.~\eqref{eq:corr}, acting on the ordered qubit pair assigned to \(T\) in Table~\ref{tab:types}.

Each channel can be decomposed into a finite set of Pauli-error realizations. For the local model, these realizations are \(\id\), \(P_a\), \(P_b\), and \(P_aP_b\), whereas for the correlated model they are \(\id\) and  \(P_aP_b\). Because the gates following the error location belong to the Clifford group, conjugating a Pauli operator by \(W_T\) produces another Pauli string:
\begin{equation}
\widetilde{E}_T
=
W_TE_TW_T^\dagger.
\label{eq:propagated_error}
\end{equation}
The Clifford propagation rules for the \(H\), \(\mathrm{CX}\), and \(\mathrm{CZ}\) gates are listed in the Supplemental Material. Any overall
phase acquired during the propagation is irrelevant at the density-matrix level because
\[
(e^{i\varphi}P)\rho(e^{i\varphi}P)^\dagger
=
P\rho P^\dagger.
\]

Acting on the ideal output state \(\ket{++}_{01}\ket{\psi}_2\), the propagated Pauli string can be factorized as
\[
\widetilde{E}_T
=
Q_{01}(T,E_T)\otimes Q_B(T,E_T),
\]
where \(Q_B(T,E_T)\in\{\id,X,Y,Z\}\) is the Pauli operator acting on Bob's qubit. After tracing out qubits \(0\) and \(1\), the corresponding error realization gives
\begin{equation}
\operatorname{Tr}_{01}
\left[
\widetilde{E}_T
\left(
\ket{++}\!\bra{++}_{01}\otimes\rho_2
\right)
\widetilde{E}_T^\dagger
\right]
=
Q_B(T,E_T)\rho_2Q_B^\dagger(T,E_T).
\label{eq:effective_mapping}
\end{equation}

By summing the probabilities of all error realizations that produce the same operator on Bob, we obtain the effective Pauli channel
\begin{equation}
\Lambda_B^{(T)}(\rho_2)
=
w_I^{(T)}\rho_2
+
w_X^{(T)}X\rho_2X
+
w_Y^{(T)}Y\rho_2Y
+
w_Z^{(T)}Z\rho_2Z,
\label{eq:effective_paulichannel}
\end{equation}
where \(w_Q^{(T)}\) is the total probability of the error realizations that induce \(Q\) on Bob, and
\[
\sum_{Q\in\{I,X,Y,Z\}}w_Q^{(T)}=1.
\]

For the stochastic Pauli models considered here, with the same Pauli axis and the probability convention defined above, the effective channels satisfy
\begin{equation}
\Lambda_B^{(T_1)}
=
\Lambda_B^{(T_6)},
\qquad
\Lambda_B^{(T_2)}
=
\Lambda_B^{(T_5)},
\qquad
\Lambda_B^{(T_3)}
=
\Lambda_B^{(T_4)}.
\label{eq:channel_symmetries}
\end{equation}
These are channel-level equivalences within the specified stochastic Pauli event sets: the paired configurations produce the same reduced state on Bob for every input state, not only the same average fidelity. They are not claimed for arbitrary coherent errors.

Using Eq.~\eqref{eq:sixaverage}, the average fidelity associated with the effective channel is
\begin{equation}
\overline{F}^{(T)}
=
w_I^{(T)}
+
\frac{
w_X^{(T)}
+
w_Y^{(T)}
+
w_Z^{(T)}
}{3}
=
1-\frac{2}{3}
\left(
1-w_I^{(T)}
\right).
\label{eq:final_avg_fid}
\end{equation}
Equation~\eqref{eq:final_avg_fid} shows that the average fidelity depends only on the total probability of nonidentity errors, \(1-w_I^{(T)}\). The state-resolved fidelities also depend on the directional distribution of the Pauli errors:
\begin{equation}
F_{\pm x}=w_I+w_X,
\qquad
F_{\pm y}=w_I+w_Y,
\qquad
F_{\pm z}=w_I+w_Z.
\label{eq:axisfidelities}
\end{equation}
Thus, the weights \(w_X\), \(w_Y\), and \(w_Z\) determine which input-state axes are preserved by the effective channel. The complete error classification, derivations of the weights, and state-resolved fidelities are provided in the Supplemental Material.

%%%%%%%%%%%%%%%%%%%%%%%%%%%%%%%%%%%%%%%%%%%%%%%%%%%%%%%%%%%%%%%%%%
\section{Results and Discussion}
\subsection{Effective Pauli maps and protected configurations}
Table~\ref{tab:mapping} summarizes the effective Pauli operators induced on Bob by the possible error realizations at the representative locations \(T_1\), \(T_2\), and \(T_3\). Under the ordered-pair correspondence defined above, the same effective mappings apply to \(T_6\), \(T_5\), and \(T_4\), respectively, as demonstrated in Sec.~\ref{sec:symmetry}. These mappings determine which error realizations affect Bob's output and provide the mechanism underlying the corresponding fidelity expressions.

At \(T_1\), the \(Z\)-axis error realizations \(Z_1\), \(Z_2\), and \(Z_1Z_2\) do not affect the state because the resource qubits are initially in \(\ket{00}_{12}\). In contrast, the correlated errors \(X_1X_2\) and \(Y_1Y_2\) produce an effective \(Y_B\) error on Bob's qubit.

At \(T_2\), a correlated same-axis Pauli error produces the corresponding Pauli error on Bob:
\[
P_1P_2\longmapsto P_B,
\qquad
P\in\{X,Y,Z\}.
\]

At \(T_3\), the Bell state \(\ket{\Phi^+}_{12}\) is preserved by the correlated operators \(X_1X_2\), \(Z_1Z_2\), and \(Y_1Y_2\), up to a global phase, a known property of same-axis Pauli correlations acting on a Bell pair \cite{Macchiavello2002}. Consequently, these error realizations induce the identity operator on Bob's reduced state. Indeed, \(X_1X_2\) and \(Z_1Z_2\) are stabilizer generators of \(\ket{\Phi^+}_{12}\), while \(Y_1Y_2=-X_1X_2Z_1Z_2\) preserves the Bell-state density operator up to an irrelevant phase. The additional result here is that the different configuration \(T_4\) induces the same reduced channel on Bob, although its errors do not stabilize the Bell resource directly.

\begin{table}[t]
\caption{Effective Pauli operator induced on Bob by single-qubit and correlated error realizations acting on the ordered pair \((a,b)\). Operators that differ only by a global phase produce the same reduced channel.}
\label{tab:mapping}
\centering
\begin{tabular}{cc ccc}
\toprule
Configuration & Axis & $P_a$ & $P_b$ & $P_a P_b$\\
\midrule
$T_1 (T_6)$ & $X$ & $Z_B$ & $X_B$ & $Y_B$\\
      & $Y$ & $Z_B$ & $X_B$ & $Y_B$\\
      & $Z$ & $\id_B$ & $\id_B$ & $\id_B$\\[1mm]
$T_2 (T_5)$ & $X$ & $\id_B$ & $X_B$ & $X_B$\\
      & $Y$ & $Z_B$ & $X_B$ & $Y_B$\\
      & $Z$ & $Z_B$ & $\id_B$ & $Z_B$\\[1mm]
$T_3 (T_4)$ & $X$ & $X_B$ & $X_B$ & $\id_B$\\
      & $Y$ & $Y_B$ & $Y_B$ & $\id_B$\\
      & $Z$ & $Z_B$ & $Z_B$ & $\id_B$\\
\bottomrule
\end{tabular}
\end{table}

The mappings in Table~\ref{tab:mapping} explain the cancellation of correlated same-axis Pauli errors at \(T_3/T_4\). At \(T_3\), the operators \(X_1X_2\), \(Y_1Y_2\), and \(Z_1Z_2\) preserve the Bell-state density operator, either exactly or up to a global phase. Therefore, this protection is not a generic consequence of entanglement: it results from the specific relation between the Bell resource and the correlated Pauli operators considered here.

At \(T_4\), the errors act on the different pair \((0,1)\) and do not directly stabilize the Bell resource. Their cancellation in Bob's output instead follows from Pauli propagation through the remaining Clifford gates and from the channel-level equivalence \(\Lambda_B^{(T_4)}=\Lambda_B^{(T_3)}\). The resulting behavior is related to stabilizer and decoherence-free-subspace ideas \cite{Zanardi1997,Lidar1998}, but here it applies to a specific resource state, circuit, and set of correlated errors rather than to a general encoded subspace.

\subsection{Analytical expressions for the average fidelity}
The analytical expressions for the three representative noise locations are summarized in Table~\ref{tab:fidelity}. The results for \(T_4\), \(T_5\), and \(T_6\) follow from the paired equivalences established in Sec.~\ref{sec:symmetry}. For convenience, we define
\begin{align}
A(p,\pstar)
&=
1-\frac{2}{3}
\left(
p+\pstar-p\pstar
\right),
\label{eq:A_function}\\
B(p,\pstar)
&=
1-\frac{2}{3}
\left(
p+\pstar-2p\pstar
\right).
\label{eq:B_function}
\end{align}
In \(A(p,\pstar)\), every realization containing at least one local error produces a nonidentity Pauli operator on Bob. The corresponding probability is \(p+\pstar-p\pstar\). In \(B(p,\pstar)\), simultaneous errors on the two qubits cancel at the output, so that only realizations in which exactly one error occurs contribute to the effective error probability \(p+\pstar-2p\pstar\). Expressions of this form and the associated nonmonotonic behavior are known in noisy-teleportation analyses \cite{Fortes2015,Fonseca2019}; their role here is to quantify the event classes obtained from the reduced-channel mapping.

\begin{table}[t]
\caption{Average teleportation fidelity \(\overline{F}\). For each ordered pair \((a,b)\) in Table~\ref{tab:types}, the single-qubit error probabilities are \(p_a=p\) and \(p_b=\pstar\).}
\label{tab:fidelity}
\centering
\small
\begin{tabular}{c c c c}
\toprule
Configuration & Axis & Local errors & Correlated errors\\
\midrule
$T_1 = T_6$ & $X, Y$ & $A(p, \pstar)$ & $1 - 2p/3$\\
            & $Z$    & $1$ & $1$\\[1mm]
$T_2 = T_5$ & $X$    & $1 - 2\pstar/3$ & $1 - 2p/3$\\
            & $Y$    & $A(p, \pstar)$ & $1 - 2p/3$\\
            & $Z$    & $1 - 2p/3$ & $1 - 2p/3$\\[1mm]
$T_3 = T_4$ & $X, Y, Z$ & $B(p, \pstar)$ & $1$\\
\bottomrule
\end{tabular}
\end{table}

\begin{figure*}[!t]
\centering
\includegraphics[width=0.98\textwidth]{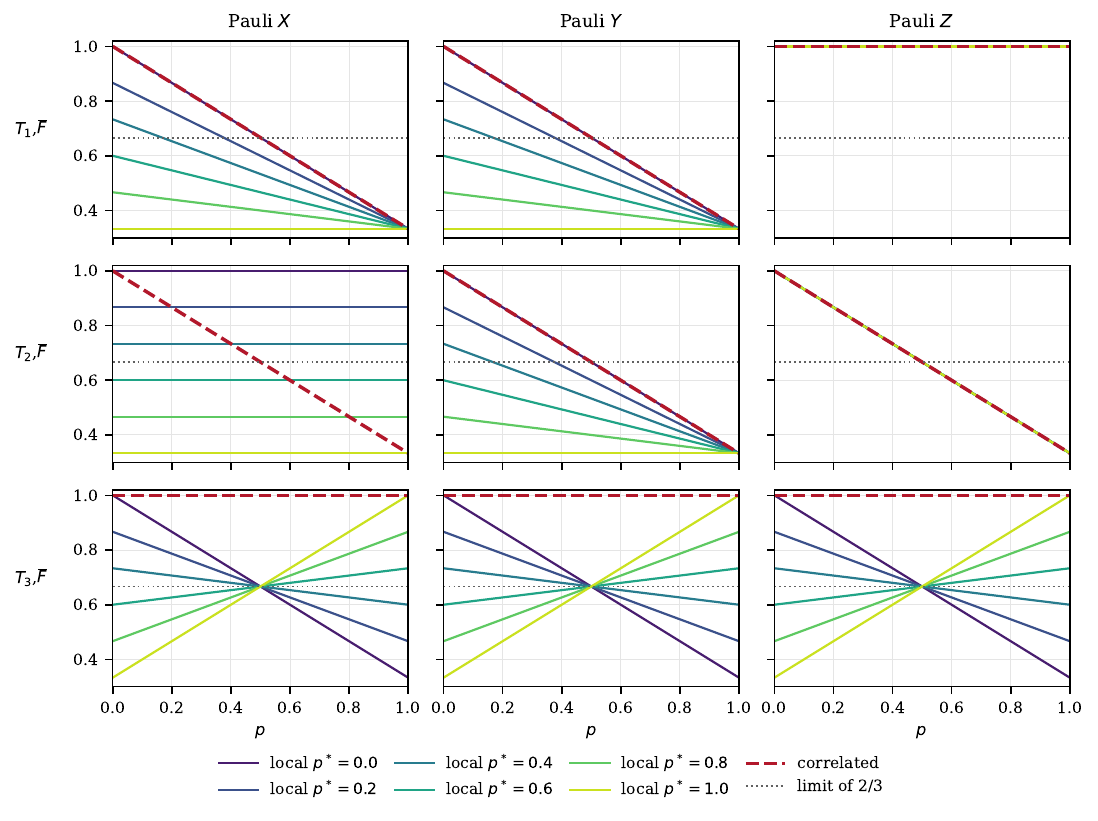}
\caption{Average teleportation fidelity as a function of the error probability \(p\) assigned to the first qubit of each ordered pair. Solid curves correspond to independent local errors for fixed \(\pstar\in\{0,0.2,0.4,0.6,0.8,1\}\); red dashed curves correspond to correlated errors, and gray dotted lines indicate the classical threshold \(2/3\). The rows show the representative locations \(T_1\), \(T_2\), and \(T_3\). The paired locations \(T_6\), \(T_5\), and \(T_4\) give the same results, respectively.}
\label{fig:fidelity-summary}
\end{figure*}

Figure~\ref{fig:fidelity-summary} illustrates the analytical expressions in Table~\ref{tab:fidelity} for \(p\in[0,1]\); it is not an independent numerical or experimental validation. The rows correspond to \(T_1\), \(T_2\), and \(T_3\), while the columns correspond to the \(X\), \(Y\), and \(Z\) axes. Solid curves show the local model for fixed \(\pstar\in\{0,0.2,0.4,0.6,0.8,1\}\), red dashed lines show the correlated model, and dotted lines mark the classical threshold \(2/3\). Comparisons between a solid and dashed curve isolate correlation only when their single-qubit marginal probabilities match, in particular along \(\pstar=p\); otherwise they also reflect different marginal rates. The paired locations \(T_6\), \(T_5\), and \(T_4\) produce the same results, respectively.

For \(T_1/T_6\), local \(Z\) errors do not affect Bob's reduced channel, because all the corresponding error realizations induce \(\id_B\). For local \(X\) or \(Y\) errors, every realization containing at least one error produces a nonidentity Pauli operator on Bob. The total nonidentity probability is
\[
1-(1-p)(1-\pstar)=p+\pstar-p\pstar,
\]
which gives the average fidelity \(A(p,\pstar)\).

In the correlated model, \(Z\) errors also produce the identity channel. Correlated \(X\) and \(Y\) errors both induce \(Y_B\) with probability \(p\), yielding \(\overline{F}=1-2p/3\). Consequently, the states \(\ket{\pm y}\) are preserved, with \(F_{\pm y}=1\), whereas the four states on the \(x\) and \(z\) axes have fidelity \(1-p\).

At \(T_2/T_5\), local \(X\) and \(Z\) errors exhibit complementary qubit
sensitivity. For \(X\) errors, an error on the first qubit of the ordered
pair induces the identity on Bob, whereas an error on the second qubit
induces \(X_B\). The effective error probability is therefore \(\pstar\),
and
\[
\overline{F}=1-\frac{2\pstar}{3}.
\]
The states \(\ket{\pm x}\) are preserved by the resulting effective
channel.

For \(Z\) errors, the roles of the two qubits are reversed. Only the error
on the first qubit affects Bob, giving
\[
\overline{F}=1-\frac{2p}{3},
\]
while the states \(\ket{0}\) and \(\ket{1}\) are preserved.

For local \(Y\) errors, the single-qubit realizations induce \(Z_B\) and \(X_B\), while the joint realization induces \(Y_B\). Thus, every realization containing at least one error contributes a nonidentity Pauli operator, giving the total probability \(p+\pstar-p\pstar\) and the average fidelity \(A(p,\pstar)\). For generic \(0<p,\pstar<1\), errors along all three Pauli axes occur in the effective channel, and no input-state axis is fully preserved. This complementary behavior identifies which physical qubit contributes to Bob's output error for a given Pauli axis.

For \(T_3/T_4\), correlated same-axis Pauli errors induce the identity operator on Bob for all three axes. Consequently, every input state is preserved and \(\overline{F}=1\). For independent local errors, a nonidentity Pauli operator reaches Bob only when exactly one of the two qubits is affected. The four possible realizations satisfy
\[
I_aI_b\mapsto I_B,\qquad
P_a\mapsto P_B,\qquad
P_b\mapsto P_B,\qquad
P_aP_b\mapsto I_B.
\]
The effective error probability is therefore
\begin{equation}
r
=
p(1-\pstar)+(1-p)\pstar
=
p+\pstar-2p\pstar.
\label{eq:xor_probability}
\end{equation}
The resulting effective channel is
\[
\Lambda_B(\rho)
=
(1-r)\rho+rP_B\rho P_B.
\]
Thus, the eigenstates of \(P_B\) are preserved even when the average fidelity is smaller than one.

The behavior of the local-error curves follows directly from the effective probability \(r\). For fixed \(\pstar<1/2\), increasing \(p\) increases the probability that exactly one error occurs, and the average fidelity decreases. For \(\pstar>1/2\), increasing \(p\) instead increases the relative weight of simultaneous errors, whose effective actions cancel on Bob, so the average fidelity increases. At \(\pstar=1/2\), \(r=1/2\) for every \(p\), giving \(\overline{F}=2/3\). This known recovery mechanism should not be interpreted as a generic improvement caused by longer or stronger environmental exposure: \(p\) and \(\pstar\) are probabilities of discrete same-axis Pauli events, and the increase follows from their specific algebraic cancellation \cite{Fortes2015}.

From Eq.~\eqref{eq:paulifid}, it is possible to see that the effective Pauli channel exceeds the classical fidelity when
\[
r<\frac{1}{2},
\]
equals it when \(r=1/2\), and falls below it when \(r>1/2\). Values below \(2/3\) remain physically valid, but low uncorrected fidelity alone does not imply irreversible loss of quantum information. In particular, when \(r>1/2\), applying one fixed operator \(P_B\) changes the residual error probability to \(1-r\), without identifying individual events. Thus the comparison with \(2/3\) refers to the uncorrected effective channel defined above.

\subsection{Equivalence of paired configurations}
\label{sec:symmetry}

The equivalence between the paired configurations follows from the propagation of the Pauli errors through the remaining Clifford gates. Consider a noise channel inserted at location \(T\), and let \(w_T(E)\) denote the probability of the error realization \(E\). After propagation through the gates following \(T\) and the partial trace over Alice's qubits, each realization induces an effective Pauli operator \(Q_B(T,E)\) on Bob. The corresponding reduced channel is
\begin{equation}
\Lambda_B^{(T)}(\rho)
=
\sum_E
w_T(E)\,
Q_B(T,E)\rho Q_B^\dagger(T,E).
\label{eq:eventchannel}
\end{equation}

Two noise configurations yield identical reduced channels on Bob whenever their error events can be paired such that corresponding realizations occur with equal probability and induce identical effective operators. Under these conditions, every term in Eq.~\eqref{eq:eventchannel} matches identically for both configurations.

For the coherent teleportation circuit studied here, Pauli propagation gives
\begin{align}
 Q_B(T_1, E) &= Q_B(T_6, E'), \nonumber \\
 Q_B(T_2, E) &= Q_B(T_5, E'), \label{eq:pairedmaps} \\
 Q_B(T_3, E) &= Q_B(T_4, E'). \nonumber
\end{align}
Here, \(E'\) is the error realization corresponding to \(E\) under the ordered-pair convention. For \(T_1,T_2,T_3\), the ordered pair is \((1,2)\), whereas for \(T_4,T_5,T_6\), it is \((0,1)\). The event correspondence is
\begin{equation}
\id\leftrightarrow\id,\qquad
P_a\leftrightarrow P_{a'},\qquad
P_b\leftrightarrow P_{b'},\qquad
P_aP_b\leftrightarrow P_{a'}P_{b'},
\label{eq:eventcorrespondence}
\end{equation}
where \((a,b)\) and \((a',b')\) are the ordered pairs associated with the two configurations.

For the local model, the four corresponding probabilities are
\begin{align}
w(\id)&=(1-p)(1-\pstar),\nonumber\\
w(P_a)&=p(1-\pstar),\nonumber\\
w(P_b)&=(1-p)\pstar,\label{eq:localeventweights}\\
w(P_aP_b)&=p\pstar.\nonumber
\end{align}

For the correlated model, the two probabilities are
\begin{equation}
w(\id)=1-p,
\qquad
w(P_aP_b)=p.
\label{eq:correlatedeventweights}
\end{equation}

Because the same probability convention is used for both ordered pairs, corresponding realizations have the same weights. Together with Eq.~\eqref{eq:pairedmaps}, this gives
\begin{equation}
\Lambda_B^{(T_1)}
=
\Lambda_B^{(T_6)},
\qquad
\Lambda_B^{(T_2)}
=
\Lambda_B^{(T_5)},
\qquad
\Lambda_B^{(T_3)}
=
\Lambda_B^{(T_4)}.
\label{eq:channelpairs}
\end{equation}

These equalities hold at the level of the reduced channel on Bob for the stated stochastic Pauli realizations and matching ordered-pair probabilities. Therefore, paired configurations produce the same output state for every input state and, consequently, the same average and state-resolved fidelities. The equivalence is stronger than an equality observed only after input-state averaging, but it is not asserted for arbitrary coherent rotations or for fixed physical-qubit rates that do not obey the relabeling convention.

\section{Conclusion and Perspectives}

In this work, we characterized the effects of statistically independent local and fully correlated single-axis Pauli errors inserted at six locations of a coherent quantum teleportation circuit. Our results show that the performance of the protocol is determined not only by the inserted errors, but by the effective channel induced on Bob's subsystem after the complete circuit execution. By propagating each Pauli error through the remaining Clifford gates, we derived analytical expressions for the average and state-resolved fidelities and identified three characteristic responses: \(Z\) errors do not affect Bob's output at \(T_1/T_6\), local \(X\) and \(Z\) errors exhibit complementary qubit sensitivity at \(T_2/T_5\), and correlated same-axis Pauli errors cancel at \(T_3/T_4\).

Furthermore, the six insertion locations form the channel-equivalent pairs \(T_1\leftrightarrow T_6\), \(T_2\leftrightarrow T_5\), and \(T_3\leftrightarrow T_4\), because corresponding error realizations occur with the same probabilities and induce the same effective Pauli operators on Bob. This equivalence holds at the level of Bob's reduced channel for every input state, rather than only after averaging the fidelity over the Bloch sphere. The identification of these channel-equivalent pairs is a central result of the present analysis.

These findings clarify, within the error models considered here, the conditions under which correlated Pauli events preserve the teleported state. Complete immunity occurs when the joint Pauli error induces \(\id_B\) on Bob's reduced state. This is distinct from a relative benefit of correlation: at equal marginal rates, correlated errors can outperform independent errors even when their joint event induces a nonidentity operator. Pre-shared entanglement alone does not guarantee either behavior.

Independent local errors also exhibit a cancellation mechanism at \(T_3/T_4\). When the same Pauli error occurs on both qubits, their effective actions on Bob cancel. A nonidentity operator therefore reaches Bob only when exactly one of the two local errors occurs, giving the effective error probability \(r=p+\pstar-2p\pstar\). For \(\pstar>1/2\), increasing \(p\) increases the probability of simultaneous errors and can therefore increase the average fidelity. This behavior does not represent a generic improvement caused by stronger physical noise: \(p\) and \(\pstar\) are probabilities of discrete Pauli errors, and the increase results from their specific cancellation in the effective channel.

The state-resolved fidelities further show that a reduction in the average fidelity does not imply uniform degradation of all input states. Eigenstates of an effective Pauli operator are preserved by that component of the channel. The event tables provide a systematic connection between error configurations and their observable effects on Bob's state, and identify components that could be prioritized in a future circuit-aware mitigation study. No specific mitigation protocol or hardware advantage is demonstrated here.

The present formulation focuses on single-axis Pauli errors, ideal Clifford gates apart from the selected noise location, and either independent local or fully correlated errors. A natural extension is to introduce a continuously tunable correlation parameter, allowing an interpolation between the local and fully correlated models. Temporal memory effects may be investigated separately. Multi-axis Pauli channels and more general two-qubit errors may also be considered to determine whether the channel equivalences and cancellation mechanisms identified here persist beyond the present error model. Finally, implementations on quantum processors may assess how these effects are modified by native multi-qubit gate errors, crosstalk, and readout errors.

\backmatter

\bmhead{Supplementary information}

Supplementary Information accompanies this article and contains the complete Pauli-event tables, representative propagation steps, derivations of the analytical fidelity expressions, and state-resolved fidelities.

\section*{Declarations}

\noindent\textbf{Funding}
L.A. acknowledges support from the Coordenação de Aperfeiçoamento de Pessoal de Nível Superior -- Brasil (CAPES) -- Finance Code 001.
J.A.d.L. acknowledges support from SECIHTI (graduate scholarship and project CBF-2025-I-1548) and from UNAM-PAPIIT (project IG101324).

\noindent\textbf{Competing interests} The authors declare no competing interests.

\noindent\textbf{Ethics approval and consent to participate} Not applicable.

\noindent\textbf{Consent for publication} Not applicable.

\noindent\textbf{Data availability} No datasets were generated or analyzed in the present study. All reported results are derived analytically in the article and its Supplementary Information.

\noindent\textbf{Materials availability} Not applicable.

\noindent\textbf{Code availability} The code used only to plot the analytical expressions in Fig.~\ref{fig:fidelity-summary} is available from the corresponding author on reasonable request.

\noindent\textbf{Author contributions}
L.A. performed the analytical calculations, analyzed the results,
and wrote the original draft of the manuscript.
L.A., J.A.d.L., and A.F. contributed to the scientific discussions
and interpretation of the results.
A.F. guided the study's conceptual development and research direction.
All authors reviewed and commented on the manuscript.

\bibliography{Thebib}

\end{document}

\maketitle

\begin{center}
\textit{Journal: Quantum Information Processing}
\end{center}

This Supplementary Information gives the analytical details underlying the effective channels and fidelities reported in the main text. It fixes the qubit-ordering and probability conventions, lists the propagated Pauli events, derives the analytical expressions, and collects the state-resolved fidelities.

\section{Circuit and conventions}

The Hilbert-space ordering is $q_0q_1q_2$, with Alice holding $q_0$ and $q_1$ and Bob holding $q_2$. The input and ancillary state are
\begin{equation}
 \rho_{\mathrm{in}}=\ket{\psi}\!\bra{\psi}_0\otimes
 \ket{00}\!\bra{00}_{12},
\end{equation}
where
\begin{equation}
 \ket{\psi}=\cos\frac{\theta}{2}\ket{0}
 +e^{i\phi}\sin\frac{\theta}{2}\ket{1}.
\end{equation}
The coherent teleportation unitary is
\begin{equation}
 U_{\rm tel}=\mathrm{CZ}_{0,2}\,\mathrm{CX}_{1\to2}\,H_0
 \mathrm{CX}_{0\to1}\,\mathrm{CX}_{1\to2}\,H_1,
 \label{eqS:utel}
\end{equation}
with the rightmost gate applied first. In the ideal case,
\begin{equation}
 U_{\rm tel}\ket{\psi}_0\ket{00}_{12}=\ket{++}_{01}\ket{\psi}_2.
 \label{eqS:ideal}
\end{equation}

The six insertion configurations are
{\small
\begin{align*}
T_1 &: P_{12}\to H_1\to \mathrm{CX}_{12}\to\mathrm{CX}_{01}
       \to H_0\to\mathrm{CX}_{12}\to\mathrm{CZ}_{02},\\
T_2 &: H_1\to P_{12}\to \mathrm{CX}_{12}\to\mathrm{CX}_{01}
       \to H_0\to\mathrm{CX}_{12}\to\mathrm{CZ}_{02},\\
T_3 &: H_1\to\mathrm{CX}_{12}\to P_{12}\to\mathrm{CX}_{01}
       \to H_0\to\mathrm{CX}_{12}\to\mathrm{CZ}_{02},\\
T_4 &: H_1\to\mathrm{CX}_{12}\to P_{01}\to\mathrm{CX}_{01}
       \to H_0\to\mathrm{CX}_{12}\to\mathrm{CZ}_{02},\\
T_5 &: H_1\to\mathrm{CX}_{12}\to\mathrm{CX}_{01}\to P_{01}
       \to H_0\to\mathrm{CX}_{12}\to\mathrm{CZ}_{02},\\
T_6 &: H_1\to\mathrm{CX}_{12}\to\mathrm{CX}_{01}\to H_0
       \to P_{01}\to\mathrm{CX}_{12}\to\mathrm{CZ}_{02}.
\end{align*}
}
Here $P_{ab}$ denotes a channel acting on the indicated pair rather than a single joint Pauli operation. Notice that $T_3$ and $T_4$ share a circuit location but act on different pairs. Likewise, $T_6$ is located before the coherent corrections and is not a readout channel.

For every ordered pair $(a,b)$, the local probabilities are $(p_a,p_b)=(p,\ps)$. Hence, the probabilities on $(q_1,q_2)$ are $(p,\ps)$ for $T_{1,2,3}$, and those on $(q_0,q_1)$ are also $(p,\ps)$ for $T_{4,5,6}$. This ordering preserves both the effective-event maps and their probabilities under the paired-configuration equivalences. The equalities are therefore conditional on relabeling the rates according to the ordered pair; they do not imply equality when fixed physical-qubit error rates are moved unchanged between configurations.

\section{Channel construction and reduced-state calculation}

For $P\in\{X,Y,Z\}$, the single-qubit channel is
\begin{equation}
 \cE_{P,p}(\rho)=(1-p)\rho+pP\rho P.
 \label{eqS:single}
\end{equation}
On a three-qubit state, $P$ is embedded by tensoring it with identities on the unaffected qubits. Independent local noise on an ordered pair $(a,b)$ is
\begin{align}
 \cE^{\rm loc}_{P;p_a,p_b}(\rho)
 =&\;(1-p_a)(1-p_b)\rho \nonumber\\
 &+p_a(1-p_b)P_a\rho P_a
 +(1-p_a)p_bP_b\rho P_b \nonumber\\
 &+p_ap_bP_aP_b\rho P_bP_a.
 \label{eqS:localexpand}
\end{align}
The perfectly correlated channel is
\begin{equation}
 \cE^{\rm corr}_{P,p}(\rho)=(1-p)\rho+pP_aP_b\rho P_bP_a.
 \label{eqS:correxpand}
\end{equation}

Let $V_T$ contain all ideal gates before location $T$ and $W_T$ all gates after it, so $U_{\rm tel}=W_TV_T$. The final state for a noisy channel with Kraus operators $K_k^{(T)}$ is
\begin{equation}
 \rho_f^{(T)}=\sum_k W_TK_k^{(T)}V_T\rho_{\rm in}
 V_T^\dagger K_k^{(T)\dagger}W_T^\dagger.
 \label{eqS:rhofinal}
\end{equation}
Bob's reduced state and the input-state fidelity are
\begin{equation}
 \rho_B^{(T)}=\Tr_{01}\rho_f^{(T)},\qquad
 F_\psi^{(T)}=\bra{\psi}\rho_B^{(T)}\ket{\psi}.
 \label{eqS:partial}
\end{equation}
Equations~\eqref{eqS:rhofinal} and \eqref{eqS:partial} provide a direct analytical construction of the reduced output channel for each possible Pauli error event and its probability.

\section{Pauli propagation through the Clifford circuit}

The circuit in Eq.~\eqref{eqS:utel} is Clifford: conjugating a Pauli string by any remaining circuit segment produces another Pauli string \cite{NielsenChuang}. The required single-gate rules include
\begin{align}
 HXH&=Z,& HZH&=X,& HYH&=-Y,\\
 \mathrm{CX}:\quad X_c&\mapsto X_cX_t,& Z_t&\mapsto Z_cZ_t,\\
 Z_c&\mapsto Z_c,& X_t&\mapsto X_t,\\
 \mathrm{CZ}:\quad X_a&\mapsto X_aZ_b,&X_b&\mapsto Z_aX_b,
\end{align}
and the corresponding products. Signs can be omitted because $(-P)\rho(-P)=P\rho P$.

For a possible Pauli error event $E$ inserted at $T$, define
\begin{equation}
 \widetilde E_T=W_TEW_T^\dagger.
\end{equation}
After acting on the ideal final state and tracing out $q_0q_1$, this operator induces a Pauli $Q_B(T,E)$ on Bob:
\begin{equation}
 \Tr_{01}\!\left[\widetilde E_T
 (\ket{++}\!\bra{++}_{01}\otimes\rho)
 \widetilde E_T^\dagger\right]=Q_B(T,E)\rho Q_B(T,E).
 \label{eqS:effective}
\end{equation}
Table~\ref{tabS:events} gives the complete event mapping. It is more informative than the average-fidelity curves because it identifies both cancellation and the output-error axis.

As one explicit example, consider the single error $X_1$ at $T_1$. Conjugating it successively by the gates following $T_1$ gives, up to phases,
\begin{equation}
 X_1 \xrightarrow{H_1} Z_1
 \xrightarrow{\mathrm{CX}_{1\to2}} Z_1
 \xrightarrow{\mathrm{CX}_{0\to1}} Z_0Z_1
 \xrightarrow{H_0} X_0Z_1
 \xrightarrow{\mathrm{CX}_{1\to2}} X_0Z_1
 \xrightarrow{\mathrm{CZ}_{0,2}} X_0Z_1Z_2.
 \label{eqS:T1example}
\end{equation}
At the paired configuration $T_6$, the corresponding event $X_0$ propagates only through the coherent corrections,
\begin{equation}
 X_0 \xrightarrow{\mathrm{CX}_{1\to2}} X_0
 \xrightarrow{\mathrm{CZ}_{0,2}} X_0Z_2.
 \label{eqS:T6example}
\end{equation}
When these strings act on $\ket{++}_{01}\otimes\rho_2$ and qubits $0$ and $1$ are traced out, both realizations induce $Z_2\rho_2Z_2$ on Bob. This example illustrates how different global propagated strings can define the same reduced error on Bob.

\begin{table*}[t]
\caption{Effective operator on Bob for every possible Pauli error event. $(a,b)=(1,2)$ for $T_{1,2,3}$ and $(a,b)=(0,1)$ for $T_{4,5,6}$. The lower half explicitly shows the paired correspondence}
\label{tabS:events}
\centering
\begin{tabular}{cc ccc@{\qquad}cc ccc}
\toprule
Type & axis & $P_a$ & $P_b$ & $P_aP_b$ & Type & axis & $P_a$ & $P_b$ & $P_aP_b$\\
\midrule
$T_1$ & $X$ & $Z_B$ & $X_B$ & $Y_B$ & $T_6$ & $X$ & $Z_B$ & $X_B$ & $Y_B$\\
      & $Y$ & $Z_B$ & $X_B$ & $Y_B$ &       & $Y$ & $Z_B$ & $X_B$ & $Y_B$\\
      & $Z$ & $\id_B$ & $\id_B$ & $\id_B$ &       & $Z$ & $\id_B$ & $\id_B$ & $\id_B$\\[1mm]
$T_2$ & $X$ & $\id_B$ & $X_B$ & $X_B$ & $T_5$ & $X$ & $\id_B$ & $X_B$ & $X_B$\\
      & $Y$ & $Z_B$ & $X_B$ & $Y_B$ &       & $Y$ & $Z_B$ & $X_B$ & $Y_B$\\
      & $Z$ & $Z_B$ & $\id_B$ & $Z_B$ &       & $Z$ & $Z_B$ & $\id_B$ & $Z_B$\\[1mm]
$T_3$ & $X$ & $X_B$ & $X_B$ & $\id_B$ & $T_4$ & $X$ & $X_B$ & $X_B$ & $\id_B$\\
      & $Y$ & $Y_B$ & $Y_B$ & $\id_B$ &       & $Y$ & $Y_B$ & $Y_B$ & $\id_B$\\
      & $Z$ & $Z_B$ & $Z_B$ & $\id_B$ &       & $Z$ & $Z_B$ & $Z_B$ & $\id_B$\\
\bottomrule
\end{tabular}
\end{table*}

Because paired configurations have identical event mappings and corresponding event probabilities, their full reduced channels are equal:
\begin{equation}
 \Lambda_{T_1}=\Lambda_{T_6},\qquad
 \Lambda_{T_2}=\Lambda_{T_5},\qquad
 \Lambda_{T_3}=\Lambda_{T_4}.
 \label{eqS:symmetry}
\end{equation}
Therefore, within the stochastic Pauli event sets and probability relabeling defined above, the equality holds for every input state and for any quantity calculated from Bob's output state, rather than only for the six-state average fidelity. No such equality is claimed for arbitrary coherent errors.

\section{From event probabilities to average fidelity}

For an effective Pauli channel
\begin{equation}
 \Lambda(\rho)=w_I\rho+w_XX\rho X+w_YY\rho Y+w_ZZ\rho Z,
 \label{eqS:paulichannel}
\end{equation}
with $ \sum_Qw_Q=1$, the average fidelity is
\begin{equation}
 \avgF=w_I+\frac{w_X+w_Y+w_Z}{3}
 =1-\frac{2}{3}(1-w_I).
 \label{eqS:avgweights}
\end{equation}
This follows because a nonidentity Pauli preserves its two eigenstates and maps each of the other four Pauli eigenstates to the orthogonal state. Thus only the total nonidentity weight is required for $\avgF$, whereas the individual weights determine the state-resolved fidelities.

\subsection{Derivation for $T_1$ and $T_6$}

For local $X$ or $Y$ noise, Table~\ref{tabS:events} and Eq.~\eqref{eqS:localexpand} give
\begin{align}
w_I&=(1-p)(1-\ps),&w_Z&=p(1-\ps),\\
w_X&=(1-p)\ps,&w_Y&=p\ps.
\end{align}
Hence $1-w_I=p+\ps-p\ps$ and
\begin{equation}
 \avgF^{\rm loc}_{T_1,X/Y}=\avgF^{\rm loc}_{T_6,X/Y}
 =1-\frac{2}{3}(p+\ps-p\ps).
 \label{eqS:T1local}
\end{equation}
For local $Z$ noise, every event maps to $\id_B$, so $\avgF=1$. In the correlated model, the joint $XX$ or $YY$ event maps to $Y_B$, yielding
\begin{equation}
 \avgF^{\rm corr}_{T_1,X/Y}=1-\frac{2p}{3},
 \qquad \avgF^{\rm corr}_{T_1,Z}=1.
 \label{eqS:T1corr}
\end{equation}
The same equations hold at $T_6$.

\subsection{Derivation for $T_2$ and $T_5$}

For local $X$ noise, all events containing $X_b$ map to $X_B$ and $X_a$ alone does not affect Bob's reduced state. The effective error probability is therefore $r_X=\ps$, independent of $p$. Conversely, for local $Z$ noise, $Z_a$ is the only relevant occurrence and $r_Z=p$. Thus
\begin{equation}
 \avgF^{\rm loc}_{T_2,X}=1-\frac{2\ps}{3},\qquad
 \avgF^{\rm loc}_{T_2,Z}=1-\frac{2p}{3}.
 \label{eqS:T2XZ}
\end{equation}
For local $Y$ noise, the three events containing at least one error reach Bob along different axes. Only the no-error event contributes to $w_I$, so
\begin{equation}
 \avgF^{\rm loc}_{T_2,Y}=1-\frac{2}{3}(p+\ps-p\ps).
 \label{eqS:T2Y}
\end{equation}
For correlated noise, the joint event maps to $P_B$ for all three axes:
\begin{equation}
 \avgF^{\rm corr}_{T_2,P}=1-\frac{2p}{3},
 \qquad P=X,Y,Z.
 \label{eqS:T2corr}
\end{equation}
These expressions also apply to $T_5$ under the same ordered-pair probability convention.

\subsection{Derivation for $T_3$ and $T_4$}

For any fixed axis, either single-qubit event maps to $P_B$, while the joint event maps to $\id_B$. An effective error therefore occurs only when exactly one local error occurs:
\begin{equation}
 r=p(1-\ps)+(1-p)\ps=p+\ps-2p\ps.
 \label{eqS:xor}
\end{equation}
It follows that
\begin{equation}
 \avgF^{\rm loc}_{T_3,P}=1-\frac{2}{3}(p+\ps-2p\ps),
 \qquad P=X,Y,Z.
 \label{eqS:T3local}
\end{equation}
In the correlated channel, only the identity and joint events occur, and both induce $\id_B$. Consequently,
\begin{equation}
 \avgF^{\rm corr}_{T_3,P}=1,\qquad P=X,Y,Z.
 \label{eqS:T3corr}
\end{equation}
The same holds for $T_4$. The protection is event-wise cancellation of the effective Pauli action. It is not implied by entanglement alone.

\section{State-resolved fidelities}
\begin{table*}[t]
\caption{Fidelity for the six input states, grouped by Bloch axis. Here $r_{\oplus}=p+\ps-2p\ps$}
\label{tabS:statefid}
\centering
\small
\begin{tabular}{cccccc}
\toprule
Type & model/axis & $F_{+x}=F_{-x}$ & $F_{+y}=F_{-y}$ & $F_{+z}=F_{-z}$ & protected axis\\
\midrule
$T_1=T_6$ & local $X$ or $Y$ & $1-p$ & $1-r_{\oplus}$ & $1-\ps$ & parameter dependent\\
            & correlated $X$ or $Y$ & $1-p$ & $1$ & $1-p$ & $y$\\
            & local/correlated $Z$ & $1$ & $1$ & $1$ & all\\[1mm]
$T_2=T_5$ & local $X$ & $1$ & $1-\ps$ & $1-\ps$ & $x$\\
            & local $Y$ & $1-p$ & $1-r_{\oplus}$ & $1-\ps$ & parameter dependent\\
            & local $Z$ & $1-p$ & $1-p$ & $1$ & $z$\\
            & correlated $P$ & $1-p\,[P\neq X]$ & $1-p\,[P\neq Y]$ & $1-p\,[P\neq Z]$ & axis of $P$\\[1mm]
$T_3=T_4$ & local $P$ & $1-r_{\oplus}[P\neq X]$ & $1-r_{\oplus}[P\neq Y]$ & $1-r_{\oplus}[P\neq Z]$ & axis of $P$\\
            & correlated $P$ & $1$ & $1$ & $1$ & all\\
\bottomrule
\end{tabular}
\end{table*}
For the channel in Eq.~\eqref{eqS:paulichannel}, the fidelities of the Pauli eigenstates are
\begin{align}
F_{\pm x}&=w_I+w_X=1-w_Y-w_Z,\\
F_{\pm y}&=w_I+w_Y=1-w_X-w_Z,\\
F_{\pm z}&=w_I+w_Z=1-w_X-w_Y.
\label{eqS:axisfid}
\end{align}
The $+$ and $-$ states on a given axis have the same fidelity for a unital Pauli channel. Table~\ref{tabS:statefid} collects the independent cases; the paired types have identical values.

In Table~\ref{tabS:statefid}, $[\mathcal{C}]$ is the indicator of condition $\mathcal{C}$: it equals one if the condition is true and zero otherwise. The table makes clear that an average fidelity below one does not imply equal degradation of all inputs. For example, the correlated $X/Y$ channels at $T_1/T_6$ induce $Y_B$ and therefore preserve $\ket{\pm i}$ with unit fidelity.

\bibliographystyle{unsrt}
\bibliography{Thebib}